\documentclass{osa-article}
\usepackage{ulem}
\journal{oe}
\articletype{Research Article}

\begin{document}

\title{An all-optical intrinsic atomic gradiometer with sub-20 fT/cm/$\sqrt{\rm Hz}$ sensitivity in a 22 $\mu$T earth-scale magnetic field}

\author{A. R. Perry,\authormark{1}\authormark{*} M. D. Bulatowicz,\authormark{2,3} M. Larsen,\authormark{3} T. G. Walker,\authormark{2} and R. Wyllie\authormark{1}}

\address{\authormark{1}Georgia Tech Research Institute, Atlanta, GA 30332, USA\\
\authormark{2}University of Wisconsin–Madison, Madison, Wisconsin 53706, USA\\
\authormark{3}Northrop Grumman Systems Corporation, Woodland Hills, CA, U.S.A.}

\email{\authormark{*}abigail.perry@gtri.gatech.edu}

%%%%%%%%%%%%%%%%%%% abstract %%%%%%%%%%%%%%%%
\begin{abstract}
In this work we demonstrate a high sensitivity atomic gradiometer capable of operation in  earth-field level environments.  We apply a light-pulse sequence at four times the Larmor frequency to achieve gradiometer sensitivity <20 fT/cm/$\sqrt{\rm Hz}$ at the finite field strength of 22 $\mu$T.  The experimental timing  sequence can be tuned to the field magnitude of interest.   Our  all-optical scalar gradiometer performs a differential measurement between two regions of a single vapor cell on a 4 cm baseline. Our results pave the way for extensions to operation in higher dimensions, vector sensitivity, and more advanced gradiometers.
\end{abstract}

%%%%%%%%%%%%%%%%%%%%%%%%%%  body  %%%%%%%%%%%%%%%%%%%%%%%%%%
\section{Introduction}
Optically pumped atomic magnetometers were first demonstrated over 60 years ago \cite{Bell1957, Bell1961}. 
In its initial development, it was realized that synchronous modulation of either the magnetic field or the optical pumping dynamics provided a powerful tool for measurement. 
Since then, breakthroughs in sensitivity \cite{Dang2010} have accompanied exploration of the fundamental physical limits, e.g. spin-exchange broadening suppression \cite{Happer1973, happer1977, Appelt1999}.
Enabling technologies like lasers and miniaturized low-power packaging have increased the fieldability of optical atomic magnetometers \cite{Sheng2017}. 
This has also led to applications in biomagnetic imaging \cite{Xia2006, Sander:12, Wyllie:12, Boto2017, Boto2018}, with long-term prospects in magnetic anomaly detection \cite{Sheinker2009} and magnetic navigation \cite{Canciani2016}. 
In all of these applications, a magnetic field source must be localized against a background of lower-spatial-frequency noise.
Hence the measurement of magnetic field gradients, as opposed to field magnitude, is an essential enabling tool for applications.
An enduring challenge is achieving high sensitivity and common-mode suppression for observing gradients with high fidelity in earth's field \cite{smullin_2009, Sheng2017, Limes2020, Zhang2020}. 

Recent work has investigated methods of spin-exchange suppression for use in earth-sized field by the addition of pulsed fields to modulate spin-precession \cite{KorverPRL2013,KorverPRL2015,Zhivun2019} and synchronous optical pulsing \cite{Han2017, Gerginov2017,Gerginov2020}. 
Here we present an optically-differenced, intrinsic magnetic gradiometer, an extension of the recently demonstrated \cite{Gerginov2017} synchronous light-pulse atomic magnetometer (SLAM), a modified low-duty cycle optical Bell-Bloom technique. 
In addition, we extend the technique to operate at high densities without excessive spin-exchange broadening of the magnetic resonance line.  
The method works well for earth-scale fields, as demonstrated here.  In the following, we report measurements of a high density two-zone SLAM with gradiometer sensitivity of 15 fT/cm/$\sqrt{\rm Hz}$ averaged over the range of 85 Hz to 156 Hz, and shot-noise sensitivity of less than 3 fT/cm/$\sqrt{\rm Hz}$. 
This sensor does not require modulating magnetic fields, which eliminates the potential for crosstalk between sensors when used in array applications.
The single recycled probe beam allows 
spatially distinct magnetometer regions to be subtracted optically, resulting in common mode noise suppression before electronic amplification and subsequent signal processing.

The underlying principles of our SLAM scheme can be described by an ensemble of atomic spins, polarized by a periodically pulsed, circularly-polarized pumping laser beam propagating along the axis $\hat{R}$.  Between pump pulses, the polarized atoms precess about the external magnetic field at the Larmor frequency $\omega_L=\gamma B$, where $\gamma$ is the gyromagnetic ratio. The probability of absorbing photons from the pumping laser is proportional to $(1-P_R)$, where $P_R$ is the component of the spin-polarization along the pump axis. For small deviations of the pump repetition frequency $\omega$ about $\omega_L$, there is an enhanced absorption of photons which brings the instantaneous spin-polarization closer into alignment with the light.  The net result is spin precession at the pulse repetition frequency but with a phase shift $\phi$ between the spin-precession and the clock driving the pump pulses:
\begin{equation}
P_R(t) = P_{\bot} \cos\left( \omega t + \phi \right) + P_{\parallel}.
\end{equation}
Here the component of the atomic polarization perpendicular to the external magnetic field $P_{\bot}=QT_{2} \left(1-\left( \hat B \cdot \hat{R}\right)^{2} \right)^{1/2}$, with transverse spin relaxation time  $T_{2}$ and average photon absorption rate $Q$. The phase shift is $\phi={\rm tan}^{-1}(\Delta\omega T_2)$, and the frequency detuning is $\Delta \omega=\omega-\omega_{L}$.
The parallel polarization component of the atomic spin, $P_{\parallel}$, has a longitudinal relaxation time $T_{1}$.

We use a co-propagating probe laser to detect the atomic spin-polarization.  
The probe is detuned far off the optical resonance, $\Delta_{\rm{opt}} \gg \Gamma_{3/2}$, where $\Gamma_{3/2}$ is the pressure-broadened linewidth of the $^{87}$Rb $5P_{3/2}$ state. 
Here the Faraday rotation of the light due to the spin-dependent index of refraction of the atoms is minimally perturbing \cite{magnetometerbook}. 
This is commonly done in spin-exchange relaxation-free magnetometers \cite{allred2002,kominis2003}. 
The probe beam acquires an optical polarization rotation
\begin{equation}
\theta(t)\approx N P_R(t),
\end{equation}
where $N$ is the number of atoms in the sample volume. Demodulation of the probe rotation angle at $\omega$ for small $\Delta\omega\approx 0$ yields in-phase and quadrature signals proportional to $N P_\bot \cos(\phi)$ and $N P_\bot \sin(\phi)$, respectively.

So far we have focused on a single co-propagating pump and probe lasers. In practical magnetic sensing applications, there are great advantages in background noise suppression to be gained by configuring pairs or arrays of magnetometers as gradiometers or differential magnetometers \cite{smullin_2009, Sheng2017, Limes2020, Zhang2020}. 
For this purpose, and especially since the Faraday rotation angles when the spins are aligned with the probe laser can be radians, it can be advantageous to send the probe light through two adjacent magnetometers so that the common mode rotation can be subtracted before converting to photocurrents. As illustrated in Figure \ref{fig:FigSLAMconfig}(a), another output is available which measures the probe laser polarization rotation as it goes through both a co-propagating pump region (zone 1) and then through a counter-propagating pump region (zone 2) before measurement. The total rotation in the two regions is given by
\begin{equation}
\delta\theta \approx N \left(P_{R,1}(t)-P_{R,2}(t) \right),
\end{equation}
where we have assumed identical $N$, $\omega$, and pump laser polarization in both zones. In the near resonant, small angle expansion limit where $\phi_{1,2}\ll 1$, the in-phase and quadrature demodulation of probe Faraday rotation is now sensitive to differences $N \left(P_{\bot,1}-P_{\bot,2}\right)$ and $N \left(P_{\bot,1}\phi_1-P_{\bot,2}\phi_2\right)$, respectively. Hence, the in-phase measurement is an error signal that can be used to balance the atomic polarization in the two zones, for example, by balancing the relative pump intensity. The quadrature measurement is an error signal proportional to the magnetic field gradient between the two regions when the atomic polarization is balanced.

\section{Design (experimental setup and sequence)}

\begin{figure}[ht!]
\centering\includegraphics[width=12cm]{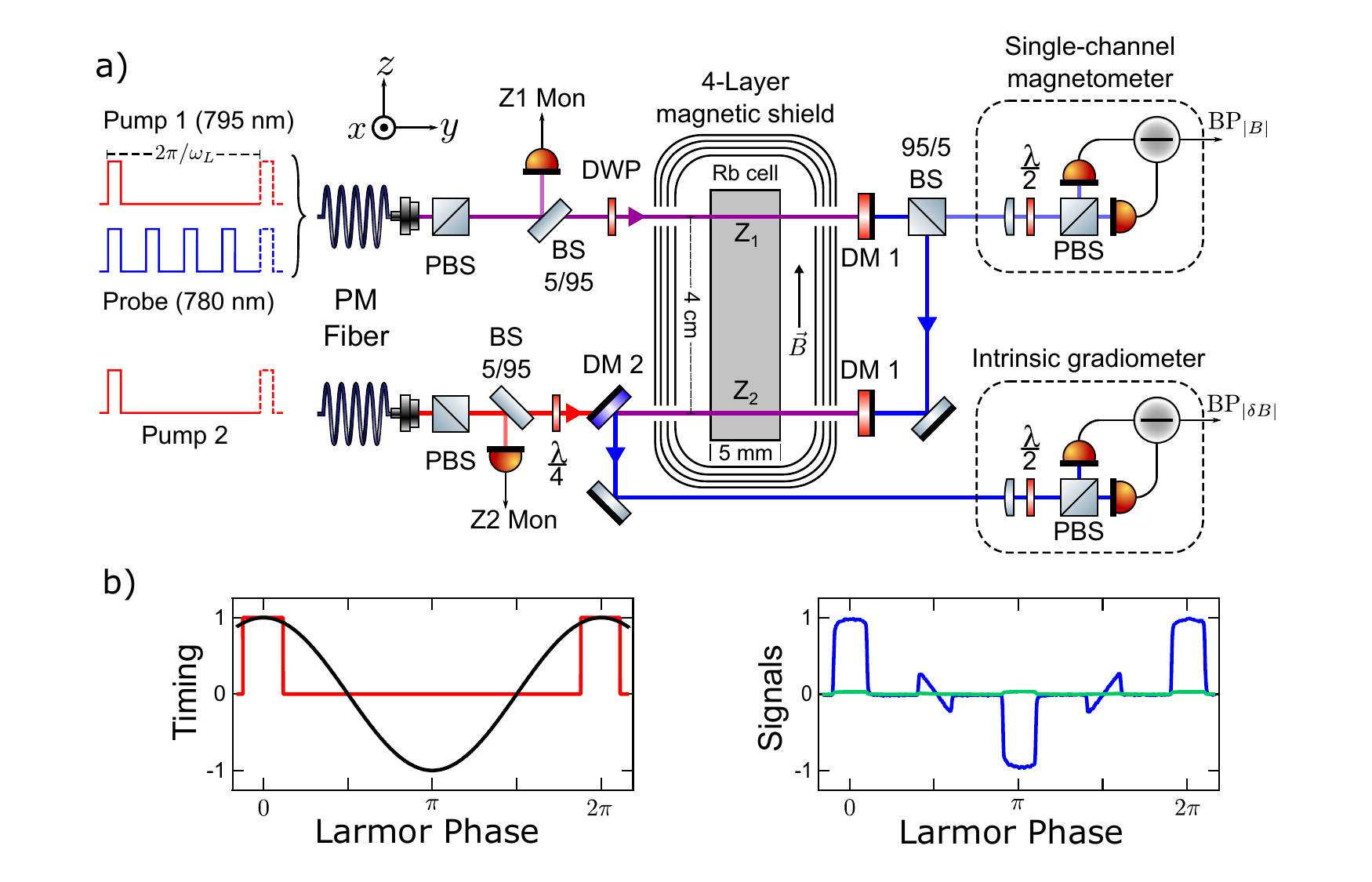}
\caption{SLAM configuration for the two-zone, intrinsic gradiometer sensor. (a) Optical layout showing the two interrogation regions using a single probe beam. Red lines indicate 795 nm pump light, blue lines indicate 780 nm probe light, and purple represents co-propagating pump and probe. Optical components include Glan-Taylor polarizers (PBS), dichroic waveplate (DWP), dichroic mirrors (DM 1, DM 2), monitor photodiodes (Z1 Mon, Z2 Mon), and beam samplers (BS). (b) The once-per-Larmor cycle pump pulse (red), the atomic polarization along the pumping direction (black), the pulsed probe sampling of the magnetometer signal $\left|B\right|$ (blue), and the pulsed probe sampling of the gradiometer signal $\left|\delta B\right|$ (green).}
\label{fig:FigSLAMconfig}
\end{figure} 

A schematic of the experimental apparatus is shown in Figure \ref{fig:FigSLAMconfig}(a). An isotopically enhanced $^{87}$Rb atomic vapor and $\approx$ 300 Torr of N$_{2}$ buffer gas is housed in a rectangular pyrex cell with inner dimensions 5 mm x 5 mm x 50 mm which is resistively heated by film heaters to an average operational temperature of $\approx$ 130 C, corresponding to a number density of $\approx 4 \cdot 10^{13}$ atoms/cm$^{3}$. There is a $\pm$10 C variance between the two zones of the cell, monitored by resistance temperature detectors.
This single cell design enables a continuously variable gradiometer baseline and guarantees common mode pressure shifts of optical resonances not easily achieveable with individually manufactured cells \cite{nardelli_conformal_2020}.
The cell, thermal management, and some steering mirrors, were placed inside a 4-layer magnetic shield with integrated coils for controlling the total field and all independent magnetic field gradient components (TwinLeaf MS-2). Two measurement zones (zone 1 and 2) spaced 4 cm along the long axis of the vapor cell have independent optical pumping beams near resonant with the 795 nm D1 line with similarly-handed circular polarization, detuned $\approx 6$ GHz below the un-pressure-shifted atomic value of the $5S_{1/2}$, $F=1$ to $5P_{1/2}$, $F=2$ transition. 
The magnetometer baseline is theoretically limited by the cross-talk free distance, which for our parameters is $<$ 500 $\mu$m \cite{PhysRevA.14.1711,Gusarov_2018,Mamishin2017}, much smaller than our experimental design considerations.
A probe beam detuned $\Delta_{\rm{opt}}\approx$ 40 GHz above the $5S_{1/2}, F=2$ to $5P_{3/2}, F=3$ transition monitors the atomic spin. All beams are collimated with a $1/e^{2}$ diameter $\approx$ 3 mm (giving a sensing volume of $\approx$ 35 mm$^{3}$). 
Accounting for attenuation due to optical coatings, Rb deposits on the cell walls, and beam clipping, we estimate the time averaged probe power to be 1.6 mW in zone 1, and 890 $\mu$W in zone 2. The time averaged pulsed pump powers are estimated to be 820 $\mu$W in zone 1, and 1.1 mW in zone 2. This results in a photodiode detection optical power of $\approx$ 80 $\mu$W in zone 1, and $\approx$ 520 $\mu$W in zone 2.
All parameters were empirically optimized by performing magnetic field sweeps to find the point of maximum sensitivity.

The probe and both pumps are first routed through acousto-optic modulators (AOMs), allowing independent timing and intensity control, discussed in further detail below. 
The probe and one pump are then combined and sent through a single polarization maintaining (PM) fiber to the experiment-side optics leading to zone 1. 
The co-propagating pump and probe beams are collimated at the fiber output and linearly polarized by a Glan-Taylor polarizer. Before entering the sensing volume, a dichroic waveplate circularly polarizes the pump while the probe remains linearly polarized. 
After the vapor cell, the pump is retro-reflected with a dichroic mirror, while the probe is transmitted. 
A pick-off mirror with balanced polarization dependence directs 5$\%$ of the transmitted probe signal to a balanced polarimeter (BP$_{|B|}$ in Figure \ref{fig:FigSLAMconfig}(a)), used to obtain the total field amplitude $|B|$. 
The probe continues to zone 2 and counter-propagates with a second pump beam of identical polarization to the first. 
Finally, the probe is separated from the pump by a dichroic mirror and its polarization detected by a second balanced polarimeter (BP$_{|\delta B|}$ in Figure \ref{fig:FigSLAMconfig}(a)), which is used to measure the magnetic field gradient. Monitor photodiodes placed in the laser beam paths of the two zones using beam samplers (BS) record the laser pulse amplitude and phases.

The left panel of Figure \ref{fig:FigSLAMconfig}(b) shows a typical timing sequence of the 795 nm pumping laser. The pump laser is pulsed at $\omega_1$, which is $\omega_1=\omega_L$ when on resonance
 (typical fields in this work are 22 $\mu$T, $\omega_{L}$ = 2$\pi\times 155$ kHz). This sets the zero phase of the Larmor precession, with a duty cycle of 10\%. The atomic spin precession, $P_{R}(t)$, depicted in the left panel of Figure \ref{fig:FigSLAMconfig}(b), is monitored by the probe laser. The probe is operated either continuously or pulsed at $4\omega_L$ and the same 10\% duty cycle as the pump. In the continuous probe case, the probe laser beam power is the cycle averaged optical power in the pulsed case. Typical raw signals using pulsed probing from both BP$_{|B|}$ and BP$_{|\delta B|}$ is shown in the right panel of Figure \ref{fig:FigSLAMconfig}(b). In the pulsed case, demodulation was achieved using a fast digitizer and differencing the average signal over the $\pi/2$ and $3\pi/2$ phases for both BP$_{|B|}$ and BP$_{|\delta B|}$. Our measurements at magnetic resonance have a zero background, allowing the use of high electronic gains on the acquired signal.

To acquire the data shown in Figures 2 and 3, experiments used a digital delay generator (SRS DG645) to switch the +80 MHz first-order AOM diffraction spot of the pump and probe beams. The optical switching, measured by a separate fast digitizer (PicoScope 5444B, 125 MHz each channel), had a rise time of $\approx$ 40 ns. 
The time series of the Larmor precession of the magnetometer and gradiometer, shown in the right panel of Figure \ref{fig:FigSLAMconfig} (b), was recorded up to 0.5 seconds. The data selected by the optical probe pulse near phases of $0$, $\pi/2$, $\pi$, $3\pi/2$ were binned and averaged to improve signal to noise. Depending on the quantity of interest, the binned averaged data can be post-processed to compute the phase, frequency, and amplitude of the atomic precession. In the Results section, the calculation of specific quantities will be described as they arise.

The data presented in Figures \ref{fig:FigLinearityCompact} and \ref{fig:FigBfieldAdev} used FPGA firmware on a NI PXIe-5171R, a fast reconfigurable oscilloscope card with available digital outputs to gate the lasers. A digital feedback loop was implemented on the FPGA to drive the $\Delta \omega$ to zero in real time and dynamically follow a changing magnetic field. This variable pulse timing can be used to calibrate the sensitivity near resonance by sweeping the pulse timing across Larmor resonance.

\section{Results}
\begin{figure}[ht!]
\centering\includegraphics[width=12cm]{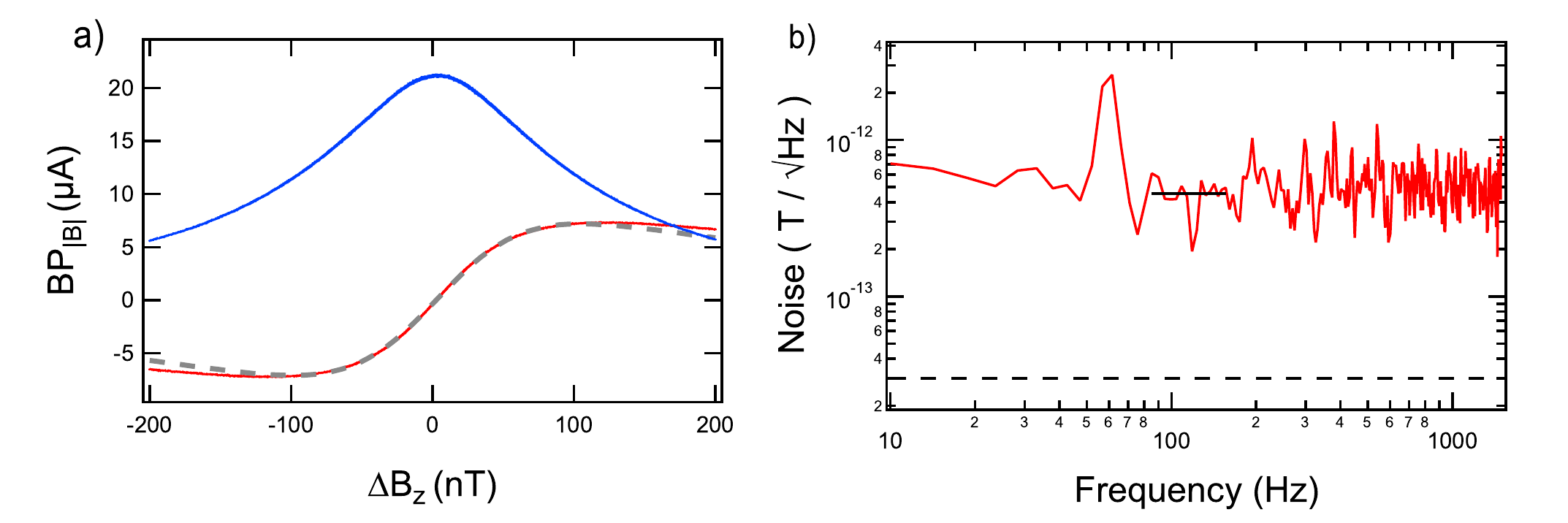}
\caption{Magnetic field calibration and sensitivity of the single channel SLAM magnetometer. The data are acquired from the balanced polarimeter signal BP$_{|B|}$ shown in the upper right of Figure \ref{fig:FigSLAMconfig}(a). (a) The external magnetic field $B_{z}$ is swept across Larmor resonance while the pump/probe pulse timing remains fixed. The blue trace shows the peaked amplitude response around Larmor resonance when the demodulation is performed using data from $0$ and $\pi$ in the Larmor cycle. The red data and dashed gray fit to the dispersive trace shows the response when the demodulation is performed using the $\pi/2$ and $3\pi/2$ data. The linear fit about the resonance gives the calibration factor between the photocurrent BP$_{|B|}$ and magnetic field. (b) Magnetometer noise floor uncorrected for frequency response is shown in red. The solid black trace is the average noise, 450 fT/$\sqrt{\rm Hz}$, over the range 85 Hz to 156 Hz. The dashed black trace shows the photon shot-noise limit. }
\label{fig:FigMagnetometer_Data}
\end{figure}

The characterization of the pulsed pump and pulsed probe mode magnetometer and gradiometer is described below. We investigated sensor operation under a continuous probe scheme, but found the noise level to be similar to the pulsed probe case although we anticipate that the pulsed mode noise is limited by timing noise, unlike the CW mode. Further investigation is necessary to determine if the technical noise limits of the CW mode are lower than the pulsed mode. Finally, we note increased coherence times and frequency resolution has been achieved by using sub-harmonic pumping, and intend to investigate these schemes in our system in the future \cite{Vasilakis2011, Shah2010a}.

To assess the sensitivity of the magnetometer, the external bias magnetic field applied along the $z$-axis (as shown in Figure \ref{fig:FigSLAMconfig}), was swept across resonance by approximately  $\pm 0.6\mu$T, and we show the demodulated signals in Figure \ref{fig:FigMagnetometer_Data}(a). 
The blue curve, which has a nominally Lorentzian lineshape, is obtained from the average of the difference between the demodulated magnetometer phase 0 and $\pi$ probe signals. 
The linewidth, given by the full width at half maximum of the resulting resonance curve, is $\approx$ 1.4 kHz. 
The red curve is obtained by demodulating the difference between the $\pi$/2 and 3$\pi$/2 probe signals, and the dashed gray trace is a fit with the functional form of the derivative of a Lorentzian, which is linear near resonance. 
For small deviations about the bias field $B_0$, the slope of the central portion of the error signal can be expressed independent of electrical gain factors by expressing the response as a ratio of the induced photocurrent from the balanced polarimeter vs the magnetic field offset from resonance, $I_{pd}/(B-B_0)$, in units of A/T. 
This helps make performance comparisons and scale noise measurements consistently across experiments, and is a key figure of merit in optimizing the magnetometer. 
This is particularly the case when it is expected that the magnetometer noise floor is dominated by the magnetic environment, rather than the fundamental limits of the sensor. 
To acquire the data shown in Figure \ref{fig:FigMagnetometer_Data}(b), we record the time trace of the atomic precession on Larmor resonance (at the steepest point of the dispersion curve in Figure \ref{fig:FigMagnetometer_Data}(a)). We then Fourier transform the precession into the frequency domain, and scale the data using the photocurrent to magnetic field calibration factor $\approx$ 140 A/T. The resulting magnetometer noise floor (between 85 Hz to 156 Hz) is 450 fT/$\sqrt{\rm Hz}$.
The magnetometer photon shot-noise limit, dashed black line in Figure \ref{fig:FigMagnetometer_Data}(b), is below 30 fT/$\sqrt{\rm Hz}$. This indicates that environmental noise sources are a limiting factor for the single-channel magnetometer.

\begin{figure}[ht!]
\centering\includegraphics[width=12cm]{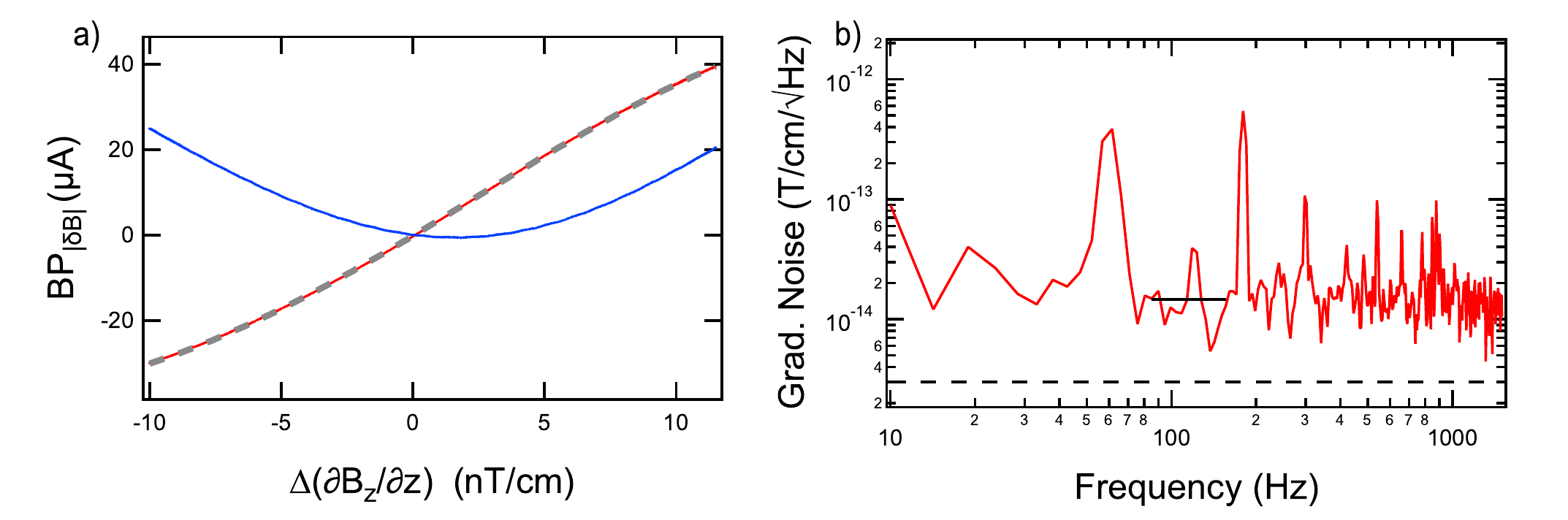}
\caption{Gradient magnetic field calibration and sensitivity of the SLAM magnetic gradiometer. The data are acquired from the balanced polarimeter signal BP$_{|\delta B|}$ shown in the lower right of Figure \ref{fig:FigSLAMconfig}(a). (a) The external magnetic gradient field $\partial_z B_z$ is swept across the zero gradient condition while the pump/probe pulse timing remains fixed.
The blue trace shows the peaked amplitude response around Larmor resonance when the demodulation is performed using data from $0$ and $\pi$ in the Larmor cycle. The red data and dashed gray fit to the dispersive trace shows the response when the demodulation is performed using the $\pi/2$ and $3\pi/2$ data. 
The linear fit about the zero gradient condition gives the calibration factor between the photocurrent BP$_{|\delta B|}$ and magnetic field.
(b) The gradiometer noise floor uncorrected for frequency response is shown in red. The solid black trace is the average noise, 15 fT/cm/$\sqrt{\rm Hz}$, over the range of 85 Hz to 156 Hz. The dashed black trace shows the photon shot-noise limit. }
\label{fig:FigGradiometer_Data}
\end{figure}

The gradiometer signal, processed similarly to the magnetometer signal, is shown in Figure \ref{fig:FigGradiometer_Data}. 
The slope of the dispersion curve in Figure \ref{fig:FigGradiometer_Data}(a) gives the gradient noise calibration factor $\approx$ 3800 A/T/cm with a magnetic gradient field sweep of approximately 15 nT/cm on either side of the magnetic resonance. 
The gradiometer noise was measured to be 15 fT/cm/$\sqrt{\rm Hz}$ for the two-zone gradiometer, shown in Figure \ref{fig:FigGradiometer_Data}(b).
The gradiometer data more nearly approaches the photon shot noise limit, <3 fT/cm/$\sqrt{\rm Hz}$, indicating good common mode rejection of environmental noise that limits the sensitivity of the single-channel magnetometer. 
The sensor also demonstrated gradient range of several nT/cm as seen in the magnetic field gradient sweep. During normal operation of the gradiometer, a small gradient is applied along the $z$-axis to bring the second zone into resonance, or balance, with the first zone. 
We found that the first zone/magnetometer response changed negligibly in sensitivity at a 14 nT/cm gradient from this two-zone balance.

\begin{figure}[ht!]
\centering\includegraphics[width=10cm]{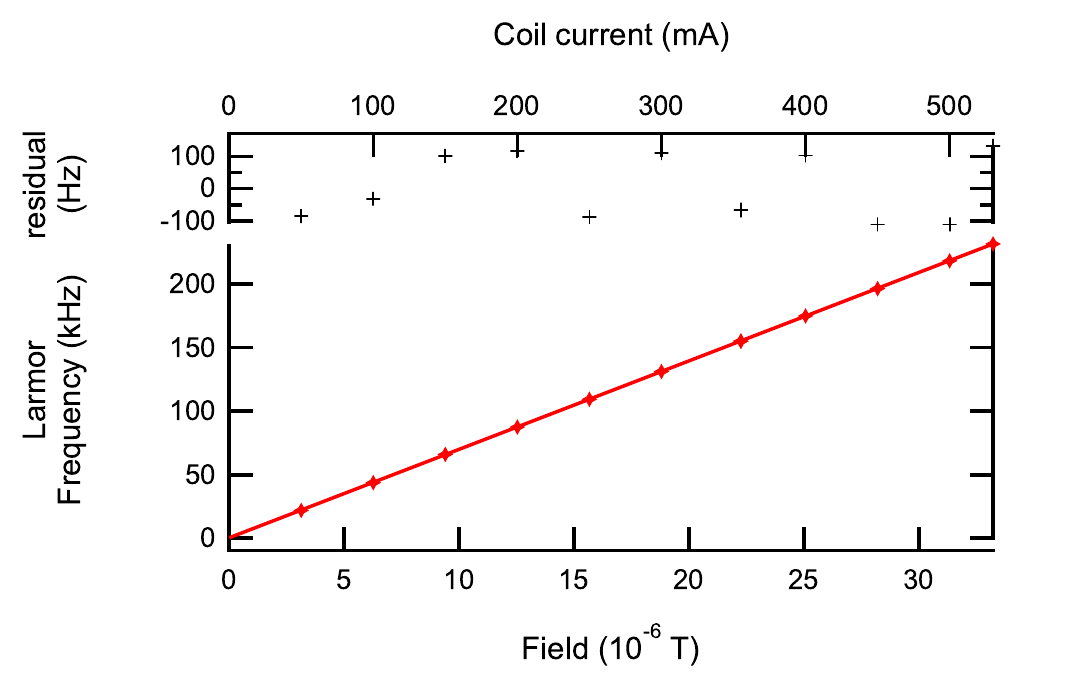}
\caption{Measurement of the self-oscillating mode locked frequency as a function of the magnetic field. The frequency response is linear from $\approx$ 3 $\mu$T to $\approx$ 30 $\mu$T.}
\label{fig:FigLinearityCompact}
\end{figure}

For sensor operation in the ambient environment, it is imperative that the sensor maintain high sensitivity while following a dynamic magnetic field environment. To this end, we have measured the timing-locked Larmor frequency as a function of applied magnetic field, shown in Figure \ref{fig:FigLinearityCompact}. The pump pulse frequency of the magnetometer remained locked as the field was varied in the range $\approx$ 3 $\mu$T to $\approx$ 30 $\mu$T. We estimate the magnitude of the quadratic Zeeman shift to be $< 50$ Hz across the range of the data presented,  within the magnitude of the fit residuals. Implementing active pulse timing feedback control enables the magnetometer to continuously track the external field in time dynamic environments.

\begin{figure}[ht!]
\centering\includegraphics[width=12cm]{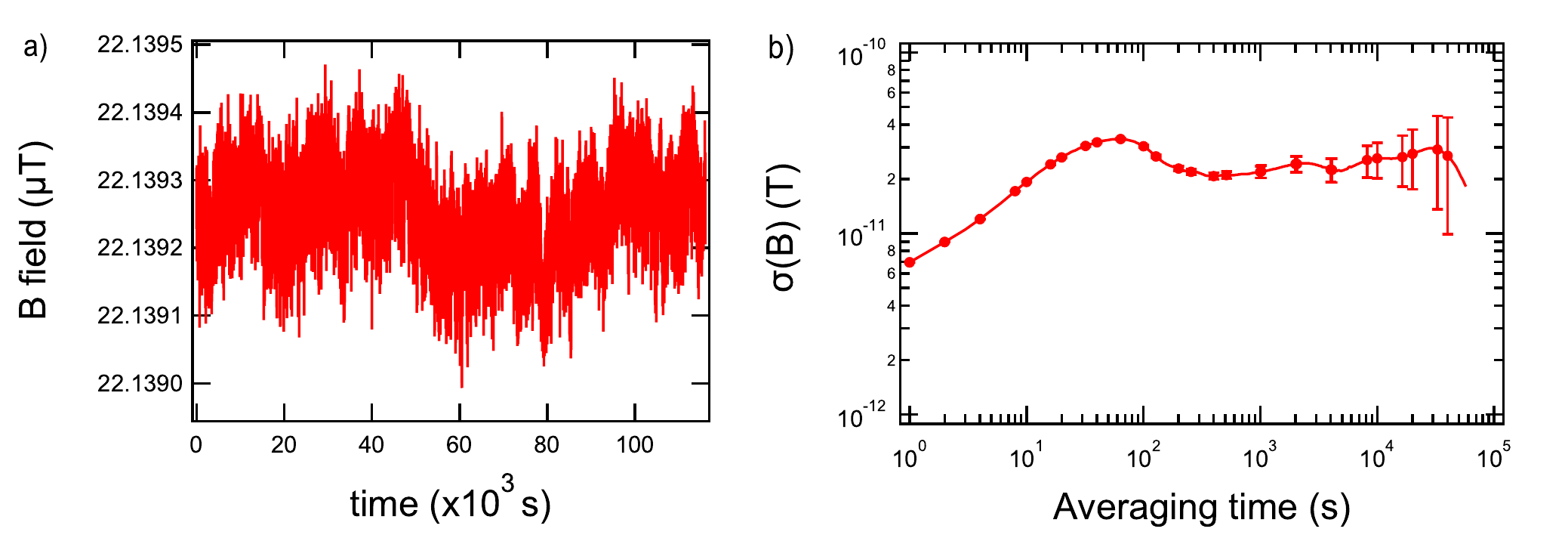}
\caption{(a) Long-term magnetic field measurement for the self-oscillating mode magnetometer, for a duration of 116000 sec. (b) The stability of the magnetometer signal as characterized by the total Allan Deviation.}
\label{fig:FigBfieldAdev}
\end{figure}

A strength of atomic sensors is that their fundamental properties are traceable to static atomic structure. 
This enables applications that depend on repeated consistent measurements, e.g. magnetic surveys. 
If the environment can be suitably controlled, then the sensor's accuracy and sensitivity should be stable at long times.  
To quantify the long term behavior of our experimental apparatus, we applied a 22 $\mu$T field using the coils external to the sensor head, and monitored the Larmor frequency of the locked SLAM magnetometer for 116000 seconds in 1 sec averaged intervals.  
This measurement is shown in Figure \ref{fig:FigBfieldAdev}(a).  Figure \ref{fig:FigBfieldAdev}(b) shows the total Allan deviation of the magnetic field (Larmor frequency) as a function of averaging time. 
The stability of the system has an upper bound below 500 pT/100,000 seconds.
We attribute the long term behavior to the experiment's sensitivity to both the oscillator timing drifts in the experiment control and to true magnetic field drifts, both from the active coils and attenuated fields from outside the shield.

\section{Limits}
Here we describe some practical limits to the SLAM measurement technique. 
The pulse length must be kept short compared to the Larmor period, $t_p\ll 2\pi/\omega_L$. The SLAM sensor low field limit is reached when the $2\pi/\omega_L\approx T2^*$, where spin-precession is too slow for coherent spin driving. In the present work this is approximately 0.6 $\mu$T.

The gradient balance depends on the local atomic polarization in each cell zone and is therefore sensitive to local pump intensity and any differential broadening mechanisms, such as probe photon absorption. While the magnetic field is measured absolutely through a frequency measurement, the gradient measurement must be calibrated.
The maximum gradient that can be measured is primarily governed by the requirement that the pump pulse frequency is the same in each zone, which limited the gradient range to $\pm$10 nT/cm as shown in Figure \ref{fig:FigGradiometer_Data}(a) (80 nT difference between zone 1 and zone 2).

All characterizations presented here were done with a magnetic field oriented perpendicular to the beam propagation axis, $\hat{R}\cdot \hat{B}=0$. The technique described here, like similar Bell-Bloom demonstrations, suffers from dead zones, where response is zero when $1-(\hat B\cdot \hat{R})^2=0$. Finally, the gradients studied here were $\partial_z B_z$. We did not investigate the dependence on gradients of other magnetic field orientations along the gradient axis.

\section{Conclusion and outlook}
Here we demonstrated a two-zone synchronous light pulsed atomic magnetometer and gradiometer capable of operating in 22 $\mu$T fields with 15 fT/cm/$\sqrt{\rm Hz}$ sensitivity.  This method is general and many simple modifications will enable increased functionality. Presently, as the angle between the external bias field and the magnetometer approaches zero the sensitivity also goes to zero, creating an angular dead zone. Implementing a multi-axis magnetometer will eliminate such dead zones enabling full 3D sensing, as well as, vector and tensor measurements in arrays. In this earliest demonstration, the pump and probe beams were recycled for use in two sensing zones.  Sensitivity may be substantially enhanced by using two sets of independently tunable pump and probe beams whose power, intensity, and timing may be independently optimized for the Larmor frequency in each sensing zone.
This operational concept simplifies the design of the apparatus at the cost of placing the burden of precision and complexity on the electronics: ongoing work is focused on optimization and characterization of this type of differential magnetometer.

\section*{Funding}
This material is based upon work supported by GTRI Internal Research and Development and the Defense Advanced Research Projects Agency (DARPA) under Contract No. 140D6318C0022. The effort depicted was sponsored by the agency set forth in the schedule of the contract, and the content of the information does not necessarily reflect the position or the policy of the Government and no official endorsement should be inferred. The views, opinions and/or findings expressed are those of the author and should not be interpreted as representing the official views or policies of the Department of Defense, Department of Interior, Interior Business Center, Acquisition Services Directorate, Division III, or the U.S. Government.

\section*{Acknowledgments}
We thank Gordon Morrison and Bob Buckley at Freedom Photonics for ongoing collaboration and discussions.

%%%%%%%%%% If using BibTeX:
\bibliography{Bibliography}

\end{document}